\documentclass[preprint]{aastex}
\usepackage{emulateapj5}
\usepackage{epsf}

\renewcommand\plotone[1]{\centerline{\epsfxsize=8.5cm\epsfbox{#1}}}
\renewcommand\plottwo[2]{\centerline{
  \epsfxsize=8.0cm\epsfbox{#1}
  \epsfxsize=8.0cm\epsfbox{#2}
}}

\slugcomment{To appear in The Astrophysical Journal}

\newcommand\x{{\bf x}}
\newcommand\y{{\bf y}}
\renewcommand\u{{\bf u}}
\renewcommand\a{\alpha}
\renewcommand\k{\kappa}
\newcommand\g{\gamma}
\newcommand\G{\Gamma}
\newcommand\bt{\beta}
\renewcommand\d{\delta}
\renewcommand\t{\theta}
\newcommand\s{\sigma}

\newcommand\refeq[1]{eq.~(\ref{eq:#1})}
\newcommand\refEq[1]{Eq.~(\ref{eq:#1})}
\newcommand\refeqs[2]{eqs.~(\ref{eq:#1}) and (\ref{eq:#2})}
\newcommand\reffig[1]{Figure~\ref{fig:#1}}

\newcommand\reftab[1]{Table~\ref{tab:#1}}

\begin{document}

\twocolumn[
\title{Analytic Cross Sections for Substructure Lensing}
\author{Charles R. Keeton\altaffilmark{1}}
\affil{
  Astronomy and Astrophysics Department,
  University of Chicago, \\
  5640 S. Ellis Ave.,
  Chicago, IL 60637
}

\begin{abstract}
The magnifications of the images in a strong gravitational lens
system are sensitive to small mass clumps in the lens potential;
this effect has been used to infer the amount of substructure in
galaxy dark matter halos.  I study the theory of substructure
lensing to identify important general features, and to compute
analytic cross sections that will facilitate further theoretical
studies.  I show that the problem of a clump anywhere along the
line of sight to a lens can be mapped onto an equivalent problem
of a clump in a simple convergence and shear field; clumps at
arbitrary redshifts are therefore not hard to handle in
calculations.  For clumps modeled as singular isothermal spheres
(SIS), I derive simple analytic estimates of the cross section for
magnification perturbations of a given strength. The results yield
two interesting conceptual points. First, lensed images with positive
parity are always made brighter by SIS clumps; images with negative
parity can be brightened but are much more likely to be dimmed.
Second, the clumps need not lie within the lens galaxy; they can be
moved in redshift by several tenths and still have a significant
lensing effect. Isolated small halos are expected to be common in
hierarchical structure formation models, but it is not yet known
whether they are abundant enough compared with clumps inside lens
galaxies to affect the interpretation of substructure lensing.
\end{abstract}

]
\altaffiltext{1}{Hubble Fellow}

\section{Introduction}

Strong gravitational lensing directly probes the mass distributions
of cosmological objects. In quadruply-imaged quasars and radio
sources, the flux ratios between the images provide a powerful
test of the smoothness of galaxy mass distributions at redshifts
$z \sim 0.2$--1.  The popular Cold Dark Matter (CDM) paradigm
predicts that galaxy dark matter halos are not smooth but rather
full of small mass clumps (e.g., Klypin et al.\ 1999; Moore et
al.\ 1999), which can perturb lens flux ratios in ways that cannot
be explained by smooth lens models (e.g., Mao \& Schneider 1998;
Metcalf \& Madau 2001; Brada\v{c} et al.\ 2002; Chiba 2002; Metcalf
\& Zhao 2002).

This ``substructure lensing'' effect has been detected both
individually (Mao \& Schneider 1998; Keeton 2001b; Brada\v{c} et
al.\ 2002) and statistically (Dalal \& Kochanek 2002a).  The
statistical detection implies that $f_{\rm clump} \sim 2\%$
(0.6--7\% at 90\% confidence) of galaxies' mass is in small clumps,
and this result has been used to constrain the power spectrum on
small scales (Dalal \& Kochanek 2002b).  It suggests that the
``missing satellite'' problem in CDM (e.g., Klypin et al.\ 1999;
Moore et al.\ 1999) is solved not by warm or exotic dark matter
(e.g., Spergel \& Steinhardt 2000; Colin, Avila-Reese \& Valenzuela
2000; Bode, Ostriker \& Turok 2001) but rather by astrophysical
mechanisms that suppress star formation in small clumps (e.g.,
Bullock, Kravtsov \& Weinberg 2000; Somerville 2002; Benson et al.\
2002). In other words, Dalal \& Kochanek (2002a) argue, real
galaxies contain numerous mass clumps that are predominantly dark.

The theory of substructure lensing is still young. To date, studies
have relied mainly on Monte Carlo simulations and have explored
relatively small parameter spaces.  To estimate the magnitude of
the effect and demonstrate its measurability. Chiba (2002) fixed
the clump population based on results from $N$-body simulations,
while Metcalf \& Madau (2001) adopted a clump mass function from
simulations and varied the clump mass fraction $f_{\rm clump}$
and the mass range. Dalal \& Kochanek (2002a) made a measurement
of $f_{\rm clump}$ for a given clump mass, interpreting it as a
weighted average over clump masses and an integral measure of
small-scale power (Dalal \& Kochanek 2002b). Further work is
needed to understand how well studies of individual clumps in
particular lens systems (e.g., Keeton 2001b) can provide a
differential measure of the clump mass function and hence
small-scale power. Even more importantly, the $f_{\rm clump}$ that
matters for lensing is the fraction of the surface mass density in
clumps, which may neither be independent of position nor represent
a global clump mass fraction.  The theory clearly needs to be
further developed and explored before the claims from substructure
lensing can be fully understood.  For that to happen, it would be
extremely helpful if we could find an analytic or semi-analytic
approach to the theory.

One important aspect of the theory has until now been neglected.
All previous studies of substructure lensing have assumed that the
clumps lie in the halos of lens galaxies, which seems reasonable
in light of known mass clumps in real galaxies (globular clusters,
satellite galaxies, etc.) and predictions from CDM. However, as
lensing is sensitive to all mass along the line of sight we should
at least consider the possibility of clumps outside the galaxy.
This issue has important bearing on the interpretation of substructure
lensing. Low-mass halos form in both CDM and WDM (Warm Dark Matter)
scenarios, but in WDM they are disrupted when they merge with larger
halos (e.g., Colin et al.\ 2000; Knebe et al.\ 2002). Thus, if the
clumps responsible for substructure lensing are in the lens galaxy
halos, they argue strongly for CDM and against WDM (Dalal \& Kochanek
2002a, 2002b). If they can be isolated, however, their ability to
rule out WDM may be weakened. Determining the relative importance
of embedded and isolated clumps will require a detailed calculation
of two effects: the number density of clumps in overdense regions
like galaxies may be much higher than the number density of
free-floating clumps (at least for CDM); but the portion of the
line of sight that pierces the lens galaxy halo is a tiny fraction
of the whole distance. That calculation is the focus of a separate
paper (J.\ Chen et al., in preparation) based on the formalism and
general results presented here.

This paper examines analytic aspects of substructure lensing, in
part to understand general features of the effect, and to obtain
simple analytic estimates of the lensing cross sections to
facilitate future theoretical studies. Section 2 focuses on the
case of a singular isothermal sphere (SIS) as a simple lens model
that is a reasonable representation of cosmological objects and yet
analytically tractable. Finch et al.\ (2002) recently gave a nice
analysis of cross sections for multiple imaging for an SIS in an
external shear. I extend their analysis by adding an external
convergence and computing cross sections of interest for substructure
lensing.  Section 3 places clumps in the context of substructure
lensing and considers for the first time the problem of clumps
along the line of sight.  I show that a clump at an arbitrary
redshift is no more difficult to handle than a clump in the halo
of the lens galaxy.  Sections 2 and 3 are independent of each other,
but Section 4 combines them to give quantitative examples for two
observed lenses. Some readers may wish at first to skip Sections 2
and 3, which are rather technical, and start with
the examples and discussion in Section 4. Finally, Section 5 offers
conclusions.

\section{SIS in an External Field}

A singular isothermal sphere, with density profile $\rho \propto r^{-2}$,
is a simple model that is often used in lensing because its simplicity
permits detailed analytic treatment (e.g., Finch et al.\ 2002) and
because it seems to be a good representation of the density profiles
of galaxies on the 5--10~kpc scales relevant for strong lensing (e.g.,
Fabbiano 1989; Kochanek 1993; Maoz \& Rix 1993; Kochanek 1996; Rix et
al.\ 1997; Treu \& Koopmans 2002).  The model has been used to represent
mass clumps for studies of substructure lensing, after taking into
account tidal stripping by the parent halo (Metcalf \& Madau 2001;
Dalal \& Kochanek 2002a).  Again, the simplicity of the SIS makes it
attractive for theoretical studies.  For the $\sim\!10^{6}\,M_\odot$
halos relevant for substructure lensing, the SIS profile does not
differ dramatically from the NFW (Navarro, Frenk \& White 1996)
profile inferred from cosmological $N$-body simulations.  If anything,
SIS halos may be somewhat less concentrated than NFW halos, leading
to slight underestimates of substructure lensing effects (see Metcalf
\& Madau 2001, especially their Figure~1).  I study the SIS model
because, given that it is at least moderately reasonable as a model,
its analytic tractability makes it extremely valuable for theoretical
studies of substructure lensing.

An SIS with velocity dispersion $\s$ produces a deflection
\begin{equation}
  \a(\x) = b\,\frac{\x}{|\x|}\,,
\end{equation}
written as a two-component vector of angles on the sky. (See
Schneider, Ehlers \& Falco 1992 for an introduction to the SIS and
lens theory in general.) The Einstein radius $b$ is given by
\begin{equation}
  b = 4\pi \left(\frac{\s}{c}\right)^2 \frac{D_{\rm ls}}{D_{\rm os}}\ ,
  \label{eq:SISb}
\end{equation}
where $D_{\rm ls}$ and $D_{\rm os}$ are angular diameter distances
between the lens and source and the observer and source,
respectively. (The distance ratio is the same for comoving
distances.) In the presence of an external convergence $\k$ and
shear $\g$, the system is described by the lens equation
\begin{eqnarray}
  \u &=& (1-\G)\x - \a(\x)\,, \label{eq:SISkg} \\
  \mbox{where}\quad
  \G &=& \left[\begin{array}{cc}
    \k+\g &   0   \\
      0   & \k-\g \\
    \end{array}\right] .
\end{eqnarray}
For convenience but without loss of generality, we are working in a
coordinate system aligned with the shear. The lensing magnification
$\mu$ is given by
\begin{equation}
  \mu^{-1} = \det\left(\frac{\partial\u}{\partial\x}\right)
  = \mu_0^{-1} - \frac{b}{r} \left[ 1 -\k - \g\cos2\t \right] ,
    \label{eq:mag}
\end{equation}
where
\begin{equation}
  \mu_0 = \det(1-\G)^{-1} = \left[(1-\k)^2-\g^2\right]^{-1}
    \label{eq:mu0}
\end{equation}
is the magnification produced by the background convergence and
shear field in the absence of the SIS.

For substructure lensing we seek to understand how the lensing
magnification is changed by the presence of the clump. In
particular, we wish to know the cross section for the fractional
difference between the magnification $\mu$ with the clump and the
magnification $\mu_0$ without it to be $\gtrsim$10--20\%. To begin,
note that there is a simple curve in the image plane with the
property that all images on the curve have magnification
$\mu = (1+\d)\mu_0$. Using \refeq{mag} we can write this curve
parametrically as
\begin{equation}
  r_\d(\t) = \frac{1+\d}{\d}\,\mu_0\,b\, \left[ 1 - \k - \g\cos2\t
    \right] . \label{eq:rd}
\end{equation}
Plugging $r_\d(\t)$ into the lens equation yields a parametric form
for the corresponding curve in the source plane,
\begin{eqnarray}
  u_\d(\t) &=& \left[ (1-\k-\g)r_\d(\t) - b \right] \cos\t\,, \nonumber \\
  v_\d(\t) &=& \left[ (1-\k+\g)r_\d(\t) - b \right] \sin\t\,. \label{eq:dcrv}
\end{eqnarray}
I refer to this as a ``$\d$-curve.'' Note that the curve is defined
only for values of $\t$ such that $r_\d(\t)>0$. Also note that $\d$
is defined using the magnification $\mu$ for a single image, not
the total magnification $\mu_{\rm tot} = \sum_i |\mu_i|$; the
difference will become important below.

The critical curve and caustic can be found by taking the limit
$\d \to \infty$. Finally, the pseudo-caustic is the curve in the
source plane that maps to the origin in the image plane (see Evans 
\& Wilkinson 1998), which can be written parametrically as
\begin{eqnarray}
  u_p(\t) &=& -b\,\cos\t\,, \nonumber \\
  v_p(\t) &=& -b\,\sin\t\,. \label{eq:cut}
\end{eqnarray}
We must now distinguish between three cases based on the values of
$\k$ and $\g$.

\subsection{Positive global parity}

The three cases are distinguished by the two eigenvalues
$\lambda_{\pm} = 1-\k\pm\g$ of the tensor $(1-\G)$. The first case
is when $\k$ and $\g$ are small enough that both eigenvalues are
positive, or $\k+\g<1$. An image produced by the external field
alone (without the SIS) would have positive parity.

\begin{figure}[t]
\plotone{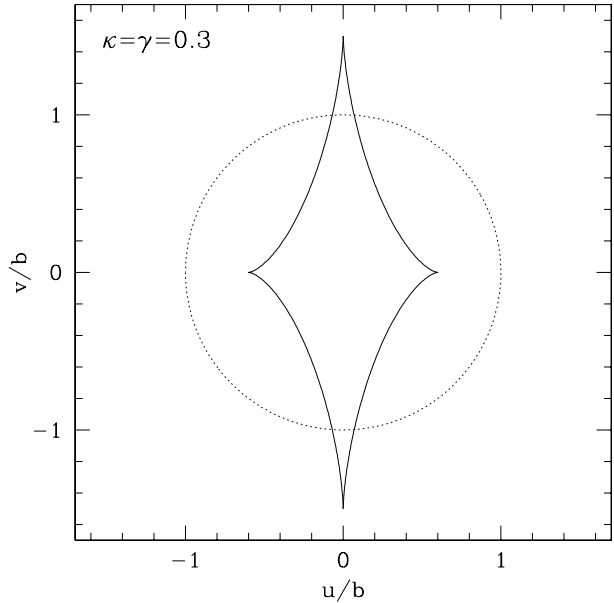}
\caption{
Caustic (solid) and pseudo-caustic (dotted) for an SIS in an external
field with $\k=\g=0.3$ and hence positive global parity.  The axes are
labeled in units of the SIS Einstein radius, $b$.
}\label{fig:causp}
\end{figure}

\reffig{causp} shows an example of the caustic and pseudo-caustic
for this case. The caustic generically has an ``astroid'' shape.
If $\g$ is small the caustic is completely enclosed by the
pseudo-caustic, and the system has standard one, two, and
four-image configurations. For larger $\g$ such that $3\g+\k>1$
(as in the example), the caustic pierces the pseudo-caustic and
produces ``naked cusps'' that correspond to an image configuration
with three bright images (e.g., Kassiola \& Kovner 1993).

\begin{figure}[t]
\plotone{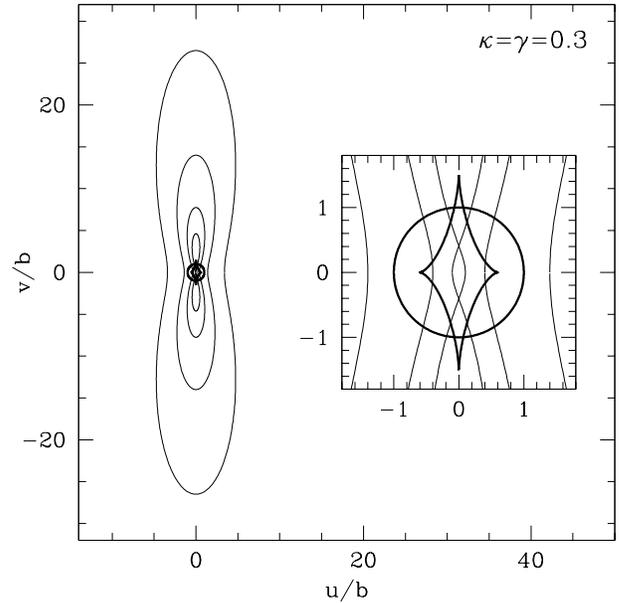}
\caption{
The light lines show $\d$-curves in the source plane, with
$\d=(0.1,0.2,0.4,0.8)$ from the outside moving in.  The heavy lines
in the middle show the caustic and pseudo-caustic from \reffig{causp}.
The inset shows a close-up of the region near the origin.
}\label{fig:deltap}
\end{figure}

\reffig{deltap} shows examples of $\d$-curves (from
eq.~\ref{eq:dcrv}). With positive global parity the curves exist
only for $\d>0$, and each one is valid for all $0 \le \t \le 2\pi$.
In general the curves have a two-lobed shape that extends far along
the vertical axis. There is a very large region where the
magnification is 10--20\% larger than the background value
($\d=0.1$--0.2), which is more than an order of magnitude larger
than the caustics.

The area enclosed by a $\d$-curve can be computed from the
parametric form as
\begin{equation}
  A(\d;\k,\g) = \int_{0}^{2\pi} \frac{1}{2} \left[ \u_\d(\t) \times
    \frac{d\u_\d}{d\t} \right]\, d\t\,, \\
\end{equation}
where $d\u_\d/d\t$ is the parametric derivative, and $\times$
indicates the vector cross product. Geometrically, the integrand is
the area of the triangle defined by the origin and the points
$\u_\d(\t)$ and $\u_\d(\t+d\t)$, and the integral simply sums all
of the triangles. The integral can be evaluated analytically,
\begin{equation}
  A(\d;\k,\g) = \frac{\pi b^2}{2\d^2}\,
    \frac{ 2(1-\k)^2 + (1-3\d^2)\g^2 }{ (1-\k)^2-\g^2 }\ .
    \label{eq:Apos}
\end{equation}
Incidentally, in the limit $\d \to \infty$ this analysis gives the
area inside the caustic as $(3/2) \pi b^2 \g^2 \mu_0$, which is a
trivial generalization to $\k \ne 0$ of the result given by Finch
et al.\ (2002).

Consider the $\d$-curve for some given value $\d$. The area within
the curve is filled by other $\d$-curves with $\d'>\d$. Thus, the
curve encloses a region where the magnification perturbation
$\mu/\mu_0-1$ is at least as large as $\d$. Consequently,
$A(\d;\k,\g)$ is a lower limit on the cross section for a
magnification perturbation stronger than $\d$. To understand why it
is only a lower limit, recall that $\d=\mu/\mu_0-1$ is defined
using the magnification $\mu$ for a single image (the image lying
on the curve in the image plane defined by eq.~\ref{eq:rd})
--- not the total magnification $\mu_{\rm tot} = \sum_i |\mu_i|$.
Since $\mu_{\rm tot} \ge \mu$, we have $\d_{\rm tot} \ge \d$. This
detail is unimportant if the $\d$-curve lies entirely outside the
caustic and pseudo-caustic, which is true for
\begin{equation}
  \d < \frac{ 1-\k-\g }{ 1-\k+3\g }\ , \label{eq:d1pos}
\end{equation}
because then all points on the $\d$-curve are singly-imaged so
$\mu_{\rm tot} = \mu$ and $\d_{\rm tot} = \d$. However, if
\refeq{d1pos} is violated then there is a region outside the
$\d$-curve but inside the caustic and/or pseudo-caustic where
$\d_{\rm tot} > \d$ (see the inset of \reffig{deltap}), so the
cross section for a {\it total\/} magnification perturbation
stronger than $\d$ is somewhat larger than $A(\d;\k,\g)$.
Nevertheless, in many cases the $\d$-curve is so much larger than
the caustic and pseudo-caustic that the additional area is
negligible, so $A(\d;\k,\g)$ should still be a good approximation
to the desired cross section.

\begin{figure}[t]
\plotone{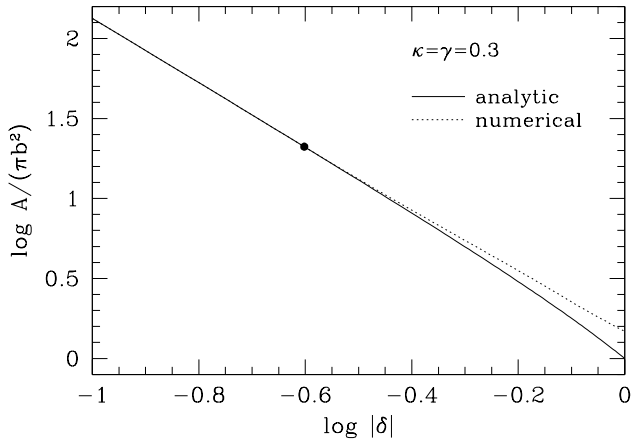}
\caption{
Estimates of the cross section for a magnification perturbation
stronger than $\d$, in units of the area $\pi b^2$ within the Einstein
ring.  The dotted curve shows the exact value computed by numerically
solving the lens equation (see text), while the solid curve shows the
analytic estimate from \refeq{Apos}.  The analytic result is exact to
the left of the dot (see eq.~\ref{eq:d1pos}).
}\label{fig:simp}
\end{figure}

To test the approximation, I compute the exact cross section
numerically. For a given source I solve the lens equation
numerically (using the algorithm and software from Keeton 2001a) to
find all of the images and compute the total magnification. I use
this process to map the magnification as a function of source
position, and then compute the cross section. \reffig{simp}
compares the exact numerical cross section to the analytic
estimate. When \refeq{d1pos} is satisfied the analytic result is
exact; when it is violated the analytic result is indeed a lower
limit on the cross section, and a very good approximation provided
the cross section is larger than a few times $\pi b^2$. Even when
the approximation is mediocre, having a simple lower limit on the
cross section is still very useful.

\subsection{Negative global parity}

In the second class of problems the eigenvalues of $(1-\G)$ have
different signs, with $1-\k-\g<0$ and $1-\k+\g>0$. An image
produced by the external field alone (without the SIS) would have
negative parity.

\begin{figure}[t]
\plotone{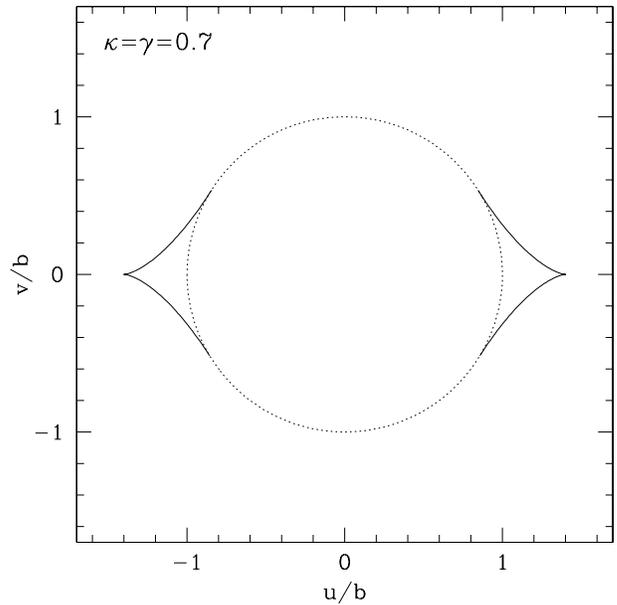}
\caption{
Caustic (solid) and pseudo-caustic (dotted) for an SIS in an external
field with $\k=\g=0.7$ and hence negative global parity.
}\label{fig:causn}
\end{figure}

\begin{figure}[t]
\plotone{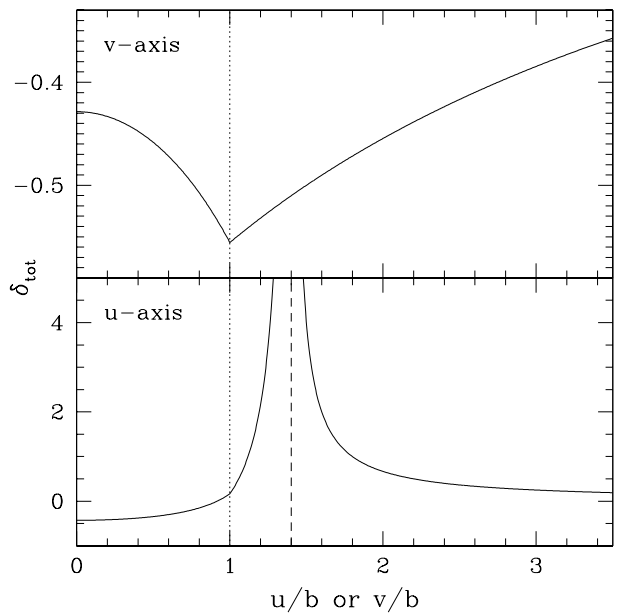}
\caption{
Total magnification perturbation $\d_{\rm tot}$ as a function of
position along the axes in \reffig{causn}.  The vertical dotted line
marks the position of the pseudo-caustic, while the vertical dashed
line marks the cusp of the caustic on the $u$-axis.
}\label{fig:dslice}
\end{figure}

\reffig{causn} shows an example of the caustic and pseudo-caustic
for this case, and \reffig{dslice} shows
$\d_{\rm tot} = |\mu_{\rm tot}/\mu_0| - 1$ as a function of
position along the axes. The caustic lies outside the
pseudo-caustic and does not close on itself but instead closes
where it touches the pseudo-caustic, at the parameter value such
that $\cos2\t=(1-\k)/\g$. A source outside both caustics produces
one image with negative parity. A source inside the pseudo-caustic
produces two images, and both have negative parity;\footnote{This
contrasts with the case of positive global parity, where a source
inside the pseudo-caustic produces one positive-parity image and
one negative-parity image.} despite having two images, the total
magnification is still smaller than the background value
($\d_{\rm tot} < 0$; see \reffig{dslice}). A source between the
caustic and pseudo-caustic produces three images, one with positive
parity and two with negative parity. The area of the pseudo-caustic
is $\pi b^2$, and the area between the pseudo-caustic and caustic
is
\begin{eqnarray}
  A_c &=& - \frac{\mu_0 b^2}{2} \biggl\{ -3(1-\k)[-(1-\k)^2+\g^2]^{1/2}
    \nonumber\\
  &&\qquad\quad + [2(1-\k)^2+\g^2]\cos^{-1}\left(\frac{1-\k}{\g}\right) \biggr\} ,
\end{eqnarray}
where $\mu_0$ is given by \refeq{mu0}, and we have $\mu_0<0$ for
negative global parity.

\begin{figure}[t]
\plotone{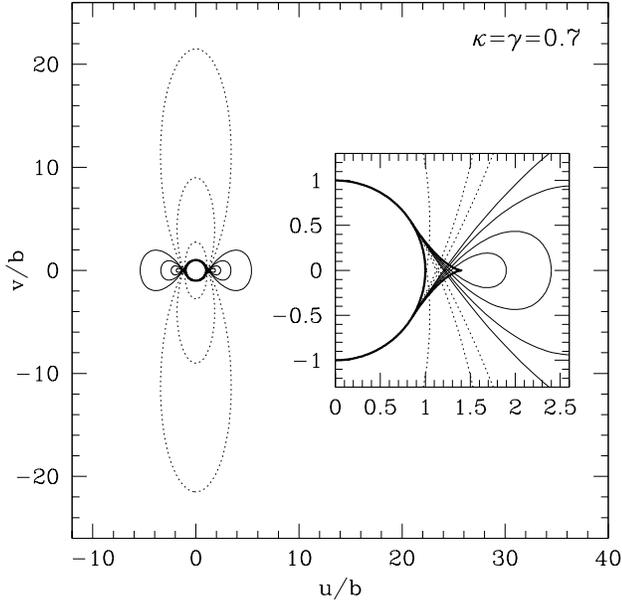}
\caption{
The light lines show $\d$-curves in the source plane; the solid
lines have $\d=(0.1,0.2,0.4,0.8)$, while the dotted lines have
$\d=(-0.1,-0.2,-0.4)$.  The heavy lines in the middle show the
caustic and pseudo-caustic from \reffig{causn}. The inset shows a
close-up of the region near the origin.
}\label{fig:deltan}
\end{figure}

\reffig{deltan} shows examples of $\d$-curves for this case. With
negative global parity the curves exist for both $\d>0$ and $\d<0$,
but each has a finite range of the parameter $\t$:
\begin{eqnarray}
  \d>0:&\ \cos2\t \ge \frac{1-\k}{\g}\ , \nonumber\\
  \d<0:&\ \cos2\t \le \frac{1-\k}{\g}\ .
\end{eqnarray}
The $\d>0$ curves have two disjoint lobes that extend to moderate
distances along the horizontal axis, while the $\d<0$ curves extend
to large distances along the vertical axis. In other words, for
negative global parity there is a moderate-sized region where the
magnification is enhanced relative to the background value, plus a
large region where the magnification is {\it reduced}.

\begin{figure}[t]
\plotone{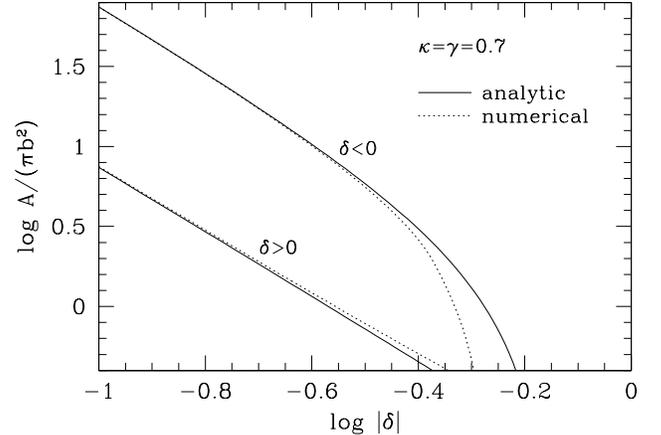}
\caption{
Estimates of the cross section for a magnification perturbation
stronger than $\d$.  Dotted curves show exact values computed by
numerically solving the lens equation (see text), while solid curves
show analytic estimate from \refeqs{Aneg1}{Aneg2}.
}\label{fig:simn}
\end{figure}

A close-up of the region near the origin shows that the
$\d$-curves intersect themselves (inside the caustic in the inset
of \reffig{deltan}); the intersection occurs at the parameter
value
\begin{equation}
  \t_\d = \sin^{-1} \left[-\frac{1-\k-\g}{2\g(1+\d)}\right]^{1/2} .
\end{equation}
Recall that $1-\k-\g<0$, so the term in square brackets is positive.
Note that this term exceeds unity when
\begin{equation}
  \d < - \frac{1-\k+\g}{2\g} < 0\,,
\end{equation}
in which case this formalism breaks down. This happens only for
large negative $\d$, at positions deep inside the pseudo-caustic
where the $\d$-curve formalism does not accurately represent the
cross section, for cases where the cross section is small and
uninteresting anyway.  So this is not a concern for most $\d<0$
cases of interest, and we can compute the enclosed area as
\begin{eqnarray}
  A_{-}(\d;\k,\g) &=& 2 \int_{\t_\d}^{\pi-\t_\d} \frac{1}{2}
    \left[ \u_\d(\t) \times \frac{d\u_\d}{d\t} \right]\, d\t\,, \nonumber\\
  &=& - \frac{ \mu_0 b^2 }{ 2\d^2 } \Bigl\{
    (\pi-2\t_\d)[2(1-\k)^2+(1-3\d^2)\g^2] \nonumber\\
  &&\qquad\qquad - f(\d;\k,\g) \Bigr\} \,,
    \label{eq:Aneg1}
\end{eqnarray}
where the integral covers one lobe, the leading factor of two
counts the other lobe, and
\begin{eqnarray}
  f(\d;\k,\g) &=& \frac{1}{1+\d}\,[1-\k+\g(1+2\d)]^{1/2}(-1+\k+\g)^{1/2}
    \nonumber \\
  && \quad\times\, [\g\d(1+3\d)-(3+\d)(1-\k)] . \label{eq:fdef}
\end{eqnarray}
\refEq{Aneg1} is the area where the magnification perturbation for
one image is stronger than $\d$ (by which I mean
$|\mu/\mu_0|-1 < \d$ since $\d<0$). It is only an approximation to
the area where the total magnification perturbation is stronger
than $\d$, because of the difference between $\d$ and $\d_{\rm
tot}$ inside the caustic and pseudo-caustic. However, \reffig{simn}
compares the analytic result with an exact numerical evaluation of
the cross section (see \S2.1 for details) and shows that the
approximation is very good provided the cross section is larger
than a few times $\pi b^2$.

A $\d$-curve with $\d>0$ always has two disjoint lobes with total area
\begin{eqnarray}
  A_{+}(\d;\k,\g) &=& 2 \int_{-\t_\d}^{\t_\d} \frac{1}{2}
    \left[ \u_\d(\t) \times \frac{d\u_\d}{d\t} \right]\, d\t\,, \nonumber\\
  &=& - \frac{ \mu_0 b^2 }{ 2\d^2 } \Bigl\{
    2\t_\d [2(1-\k)^2+(1-3\d^2)\g^2] \nonumber\\
  &&\qquad\qquad + f(\d;\k,\g) \Bigr\} \,,
    \label{eq:Aneg2}
\end{eqnarray}
where again the integral covers one lobe, the leading factor of two
counts the other lobe, and $f(\d;\k,\g)$ is given by \refeq{fdef}.
This is a lower limit on the area where the total magnification
perturbation is stronger than $\d$, because it omits some area
inside the caustic where $\mu_{\rm tot} > \mu$ and $\d_{\rm tot} > \d$.
\reffig{simn} shows, though, that the analytic result is a very good
approximation to the exact numerical cross section.

\subsection{Doubly negative global parity}

In the final class of problems the eigenvalues of $(1-\G)$ are both
negative, $1-\k-\g<0$ and $1-\k+\g<0$. In real lenses this case
occurs only for an image very near the center of the lens galaxy,
where the high surface density leads to strong demagnification.
Observations of such ``core'' images are very rare (e.g., Rusin
\& Ma 2001; Keeton 2002), so in practice this case is unimportant
for studies of substructure lensing.

\section{Substructure Lensing}

I now turn to the case of substructure lensing, in which a small
clump lies along the line of sight to one of the images in a strong
lens system. On the scale of the clump --- for realistic situations
the characteristic clump Einstein radius is $\sim$1 mas (e.g.,
Dalal \& Kochanek 2002a) --- the effect of the main lens galaxy can
be approximated as a simple convergence and shear. In this section
I manipulate the lens equation to map the galaxy+clump problem onto
the simpler problem of a clump in an external field. The analysis
in this section is general and does not assume any specific form
for either the galaxy or the clump.

\begin{figure}[t]
\plotone{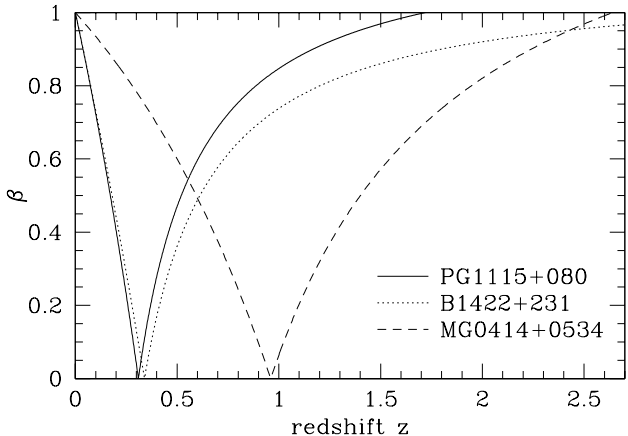}
\caption{
$\bt$ versus the clump redshift, for three lens systems:
PG~1115+080      with $z_s=1.72$ and $z_l=0.31$;
B1422+231        with $z_s=3.62$ and $z_l=0.34$;
and MG~0414+0534 with $z_s=2.64$ and $z_l=0.96$
(Weymann et al.\ 1980; Patnaik et al.\ 1992; Lawrence et al.\ 1995;
Kundi\'c et al.\ 1997a, 1997b; Tonry \& Kochanek 1999).
The cosmology is $\Omega_M=0.3$ and $\Omega_\Lambda=0.7$.
}\label{fig:beta}
\end{figure}

The clump may be embedded in the halo of the main lens galaxy, but
for generality I also consider the possibility that it lies
elsewhere along the line of sight. This requires formulating the
lens equation with two lens planes. Suppose a deflector at redshift
$z_1$ produces deflection $\a_1(\x)$, and a second deflector at
redshift $z_2 \ge z_1$ produces deflection $\a_2(\x)$. The lens
equation then has the form (e.g., Schneider et al.\ 1992)
\begin{equation}
  \u = \x - \a_1(\x) - \a_2\left[ \x - \bt \a_1(\x) \right] ,
    \label{eq:lens2}
\end{equation}
where the factor
\begin{equation}
  \bt = \frac{ D_{12} D_{\rm os} }{ D_{\rm o2} D_{\rm 1s} }
\end{equation}
encodes the redshift difference in terms of a distance
ratio,\footnote{The distance ratio is the same for angular diameter
distances or comoving distances.} where $D_{ij} = D(z_i,z_j)$ and
``o'' and ``s'' refer to the observer and source respectively.
\reffig{beta} shows that $\bt$ vanishes if the two redshifts are
the same, and monotonically approaches unity as either redshift
approaches the observer or the source. Given the form of
\refeq{lens2}, we must distinguish between cases where the clump is
in the foreground or background of the galaxy. The case of a clump
embedded in the galaxy is given by the limit $\bt \to 0$ of either
case.

\subsection{Foreground clump}

Let $\a_g(\x)$ be the deflection of the galaxy\footnote{The ``galaxy''
may be smooth, but it could also include other clumps that are far
from the clump in question. It may also include any external tidal
shear from large objects near the lens galaxy. In the language of
substructure lensing, it represents the macromodel.} without the
clump and $\a_c(\x)$ be the deflection from the clump in question.
Suppose that without the clump the galaxy takes a source at position
$\u$ and produces an image at position $\x_0$, so the unperturbed
lens equation is
\begin{equation}
  \u = \x_0 - \a_g(\x_0)\,. \label{eq:unpert}
\end{equation}
With the clump present, the lens takes the same source to an image
at $\x$ given by the perturbed lens equation (compare
eq.~\ref{eq:lens2})
\begin{equation}
  \u = \x - \a_c(\x-\x_c) - \a_g\left[\x-\bt\a_g(\x)\right] , \label{eq:pert1}
\end{equation}
for a clump in the foreground of the galaxy. Note that we write
$\a_c(\x-\x_c)$ to emphasize that the clump deflection depends only
on the position relative to the clump center $\x_c$. Subtracting
\refeqs{unpert}{pert1}, we find an equation relating $\x_0$ and
$\x$,
\begin{equation}
  0 = (\x-\x_0) - \left\{ \a_g\left[\x-\bt\a_c(\x)\right]-\a_g(\x_0)\right\}
    - \a_c(\x-\x_c)\,. \label{eq:fg1}
\end{equation}
If the clump is small relative to the galaxy such that the
perturbation $\x-\x_0$ is small relative to length scales
associated with the galaxy, then we can expand $\a_g(\x)$ in a
Taylor series and write
\begin{eqnarray}
  \a_g(\x) &=& \a_g(\x_0) + \G\cdot(\x-\x_0)\,, \label{eq:galser} \\
  \mbox{where}\quad
  \G &\equiv& \left[\begin{array}{cc}
\frac{\partial\a_{gx}}{\partial x} & \frac{\partial\a_{gx}}{\partial y} \\
\frac{\partial\a_{gy}}{\partial x} & \frac{\partial\a_{gy}}{\partial y} \\
    \end{array}\right] , \label{eq:galG}
\end{eqnarray}
where the derivatives in $\G$ are evaluated at the unperturbed
image position $\x_0$.  In the language of substructure lensing,
$\G$ represents the convergence and shear from the macromodel.
Using \refeq{galser} in \refeq{fg1} yields
\begin{equation}
  0 = (1-\G)(\x-\x_0) - (1-\bt\G)\a_c(\x-\x_c)\,. \label{eq:fg2}
\end{equation}
Now define $\y=\x-\x_c$ as a coordinate system centered on the clump
position, and multiply \refeq{fg2} on the left by $(1-\bt\G)^{-1}$
to obtain
\begin{equation}
  \u_{\rm eff} = (1-\G_{\rm eff})\y - \a_c(\y)\,, \label{eq:fgeff}
\end{equation}
where
\begin{eqnarray}
  \u_{\rm eff} &\equiv& -(1-\bt\G)^{-1} (1-\G) (\x_c-\x_0)\,, \\
  \G_{\rm eff} &\equiv& 1 - (1-\bt\G)^{-1} (1-\G)\,.
\end{eqnarray}
\refEq{fgeff} has the same form as \refeq{SISkg}, so we have mapped
the general line-of-sight clump problem onto an equivalent problem of
a clump in a simple external field, where the effective convergence
and shear are
\begin{eqnarray}
\k_{\rm eff} &=& \frac{ (1-\bt)[\k-\bt(\k^2-\g^2)] }{ (1-\bt\k)^2-(\bt\g)^2 }\ ,
  \label{eq:keff} \\
\g_{\rm eff} &=& \frac{ (1-\bt)\g }{ (1-\bt\k)^2-(\bt\g)^2 }\ .
  \label{eq:geff}
\end{eqnarray}
Note that the ratio $\mu/\mu_0$ is the same for the general
line-of-sight clump problem and the equivalent clump plus external
field problem, so $\d$ is the same for both problems. Taking
advantage of the mapping to the simpler problem, we can write the
cross section $\s$ for a magnification perturbation stronger than
$\d$ as
\begin{eqnarray}
  \s(\d;\k,\g,\bt) &=& \int f_\d\,d\x_c
  = \int f_\d\,\left|\frac{\partial\x_c}{\partial\u_{\rm eff}}\right|\,
    d\u_{\rm eff} \nonumber\\
  &=& \left|\frac{\det(1-\bt\G)}{\det(1-\G)}\right|
    A(\d;\k_{\rm eff},\g_{\rm eff})\,,
    \label{eq:sigmap}
\end{eqnarray}
where $f_\d$ is unity if the clump position $\x_c$ corresponds to
a perturbation stronger than $\d$, and zero otherwise. In other
words, the cross section for the general line-of-sight clump
problem is simply a multiplicative factor times the cross section
for the equivalent clump plus external field problem. For an SIS
clump, $A(\d;\k_{\rm eff},\g_{\rm eff})$ is given by
eq.~(\ref{eq:Apos}), (\ref{eq:Aneg1}), or (\ref{eq:Aneg2}).

It is important to remark on the parity. The parity of the equivalent
problem is given by
\begin{equation}
  1 - \k_{\rm eff} - \g_{\rm eff} = \frac{ 1-\k-\g }{ 1-\bt(\k+\g) }\ .
\end{equation}
If the global parity in the full problem is positive ($1-\k-\g>0$),
then the parity of the equivalent problem is always positive
($1-\k_{\rm eff}-\g_{\rm eff}>0$). However, if the global parity in
the full problem is negative ($1-\k-\g<1$ and $1-\k+\g>0$), then
the parity of the effective problem can be either negative or
positive depending on the value of $\bt$:
\begin{equation}
  1 - \k_{\rm eff} - \g_{\rm eff} = \cases{
    <0, & $\bt<(\k+\g)^{-1}$ \cr
    >0, & $\bt>(\k+\g)^{-1}$
  } \label{eq:effparity}
\end{equation}
This is an important detail for clumps that lie along the line of
sight but far in redshift from a lensed image with negative parity,
as we shall see in \S4.

\subsection{Background clump}

For a clump in the background of the galaxy, the perturbed lens
equation is
\begin{equation}
  \u = \x - \a_g(\x) - \a_c\left[\x-\bt\a_g(\x)-\x_c\right] . \label{eq:pert2}
\end{equation}
Again subtracting the unperturbed lens equation \refeq{unpert} and
using \refeq{galser} to approximate $\a_g(\x)$, we find the
equation relating $\x_0$ and $\x$ to be
\begin{eqnarray}
  0 &=& (1-\G)(\x-\x_0) \\
  &&\quad - \a_c\left[ (1-\bt\G)(\x-\x_0) + \x_0 - \bt\a_g(\x_0)
    - \x_c \right] . \nonumber
\end{eqnarray}
Changing variables to
\begin{equation}
  \y = (1-\bt\G)(\x-\x_0) + \x_0 - \bt\a_g(\x_0) - \x_c\,,
\end{equation}
we find
\begin{equation}
  \u_{\rm eff} = (1-\G_{\rm eff})\y - \a_c(\y)\,,
\end{equation}
where
\begin{eqnarray}
  \u_{\rm eff} &\equiv& -(1-\G) (1-\bt\G)^{-1} [\x_c-\x_0+\bt\a_g(\x_0)]\,, \\
  \G_{\rm eff} &\equiv& 1 - (1-\G) (1-\bt\G)^{-1}\,.
\end{eqnarray}
We have again mapped the general line-of-sight clump problem onto
the simple problem of a clump in an external field, where the
effective convergence and shear are again given by
\refeqs{keff}{geff}. The cross section for a magnification
perturbation stronger than $\d$ is then
\begin{eqnarray}
  \s(\d;\k,\g,\bt) &=& \int f_\d\,d\x_c
  = \int f_\d\,\left|\frac{\partial\x_c}{\partial\u_{\rm eff}}\right|\,
    d\u_{\rm eff} \nonumber\\
  &=& \left|\frac{\det(1-\bt\G)}{\det(1-\G)}\right|
    A(\d;\k_{\rm eff},\g_{\rm eff})\,,
    \label{eq:sigmap2}
\end{eqnarray}
so again the cross section for the general line-of-sight clump
problem is simply a multiplicative factor times the cross section
for the equivalent clump plus external field problem. The parity
comments from the end of \S3.1 apply here as well. Note that the
foreground and background clump cases involve different
manipulations of the lens equation but arrive at the same mapping
of the line-of-sight clump problem to the clump plus external field
problem (eqs.~\ref{eq:sigmap} and \ref{eq:sigmap2}).

\section{Sample SIS Cross Sections}

\subsection{Procedure}

If we assume SIS clumps, we can combine results from the previous
two sections to make quantitative estimates of cross sections for
substructure lensing in real lenses. The procedure is as follows.
\begin{itemize}

\item
Select a lens and get the source and lens redshifts. Select one
image and get values for $\k$ and $\g$ from a smooth macromodel.

\item
Choose a perturbation threshold $\d$, guided by data. For example,
if the observed flux is 20\% fainter than predicted by smooth
models then one would set $\d=-0.2$ to estimate the probability
that substructure produces an effect at least as strong as
observed. Alternatively, if all other statistical and systematic
effects are thought to be smaller than 20\% then one could set
$\d=\pm0.2$ to estimate the probability of a noticeable
substructure effect.

\item
Pick a redshift for the clump and compute $\bt$.

\item
Use \refeqs{keff}{geff} to compute $\k_{\rm eff}$ and
$\g_{\rm eff}$, mapping the original problem to the equivalent
problem of a clump in an external field. Compute the effective
cross section using results from \S2, where the parity and sign of
$\d$ determine which formula is appropriate:
\begin{eqnarray}
  && 1 - \k_{\rm eff} - \g_{\rm eff} > 0 : \nonumber\\
  &&\quad A(\d;\k_{\rm eff},\g_{\rm eff}) = \cases{
    \mbox{\refeq{Apos}} & $\d>0$ \cr
    0                   & $\d<0$ \cr
  } \\
  && 1 - \k_{\rm eff} - \g_{\rm eff} < 0 : \nonumber\\
  &&\quad A(\d;\k_{\rm eff},\g_{\rm eff}) = \cases{
    \mbox{\refeq{Aneg1}} & $\d<0$ \cr
    \mbox{\refeq{Aneg2}} & $\d>0$ \cr
  }
\end{eqnarray}

\item
Use \refeq{sigmap} to convert back to the cross section for the
original problem.

\end{itemize}

\subsection{Data}

I consider two systems that seem to require substructure lensing.
B1422+231 is a radio-loud quasar at redshift $z_s=3.62$ lensed into
four images by an early-type galaxies in a poor group of galaxies
at redshift $z_l=0.34$ (Patnaik et al.\ 1992; Kundi\'c et al.\
1997b; Tonry 1998). Images A and C are bright positive-parity
images, while B is a bright negative-parity image and D is a faint
negative-parity image. In the geometric language used by Schechter
\& Wambsganss (2002), A and C lie at minima of the time delay
surface, while B and D lie at saddlepoints.\footnote{There is
presumably a faint fifth image lying at a maximum of the time delay
surface near the center of the lens galaxy, which would have doubly
negative parity, but if it exists it is too faint to have been
observed (Patnaik et al.\ 1992).} \reftab{macro} gives the
convergence and shear from macromodels of the system. The image
positions can be fit to high precision, but the flux ratios have
long been a problem (e.g., Hogg \& Blandford 1994; Kormann,
Schneider \& Bartelmann 1994; Keeton, Kochanek \& Seljak 1997;
Mao \& Schneider 1998; Brada\v{c} et al.\ 2002; Metcalf \& Zhao
2002).  Keeton (2001b) argues that there must be a clump lying in
front of image A with mass $10^{5-6}\,M_\odot$ if it is a point
mass, or $10^{6-7}\,M_\odot$ if it is an SIS.

\begin{deluxetable}{ccrrr}
\tablewidth{230pt}
\tablehead{
 \colhead{Lens} & \colhead{Image} & \colhead{$\k$} & \colhead{$\g$} & \colhead{$\mu_0$}
}
\startdata
B1422+231
& A     & $0.384$ & $0.476$ & $  6.57$ \\
& B     & $0.471$ & $0.634$ & $ -8.26$ \\
& C     & $0.364$ & $0.414$ & $  4.29$ \\
& D     & $1.863$ & $2.025$ & $ -0.30$ \\
\tableline
PG~1115+080
& A$_1$ & $0.532$ & $0.412$ & $ 19.96$ \\
& A$_2$ & $0.551$ & $0.504$ & $-19.10$ \\
& B     & $0.663$ & $0.644$ & $ -3.32$ \\
& C     & $0.469$ & $0.286$ & $  5.00$ \\
\enddata
\tablecomments{
Convergences, shears, and magnifications from macromodels.
B1422+231 is modeled as a singular isothermal ellipsoid (SIE) plus
an external shear (Keeton 2001b), while PG~1115+080 is modeled as
an SIE plus an additional SIS representing the surrounding poor
group of galaxies (Impey et al.\ 1998).
}\label{tab:macro}
\end{deluxetable}

PG~1115+080 is a radio-quiet quasar at $z_s=1.72$ lensed into four
images by an early-type galaxy in a poor group of galaxies at
redshift $z_l=0.31$ (Weymann et al.\ 1980; Kundi\'c et al.\ 1997a;
Tonry 1998). Images A$_1$ and C are positive-parity images (minima),
while A$_2$ and B are negative-parity images (saddlepoints). Again
the image positions can be fit to high precision, but the observed
A$_2$/A$_1$ flux ratio of $\sim$0.65 is hard to reconcile with a
generic prediction of $\sim0.96$ for smooth macromodels (e.g.,
Impey et al.\ 1998; Metcalf \& Zhao 2002).

\subsection{Results}

\begin{figure*}[t]
\plottwo{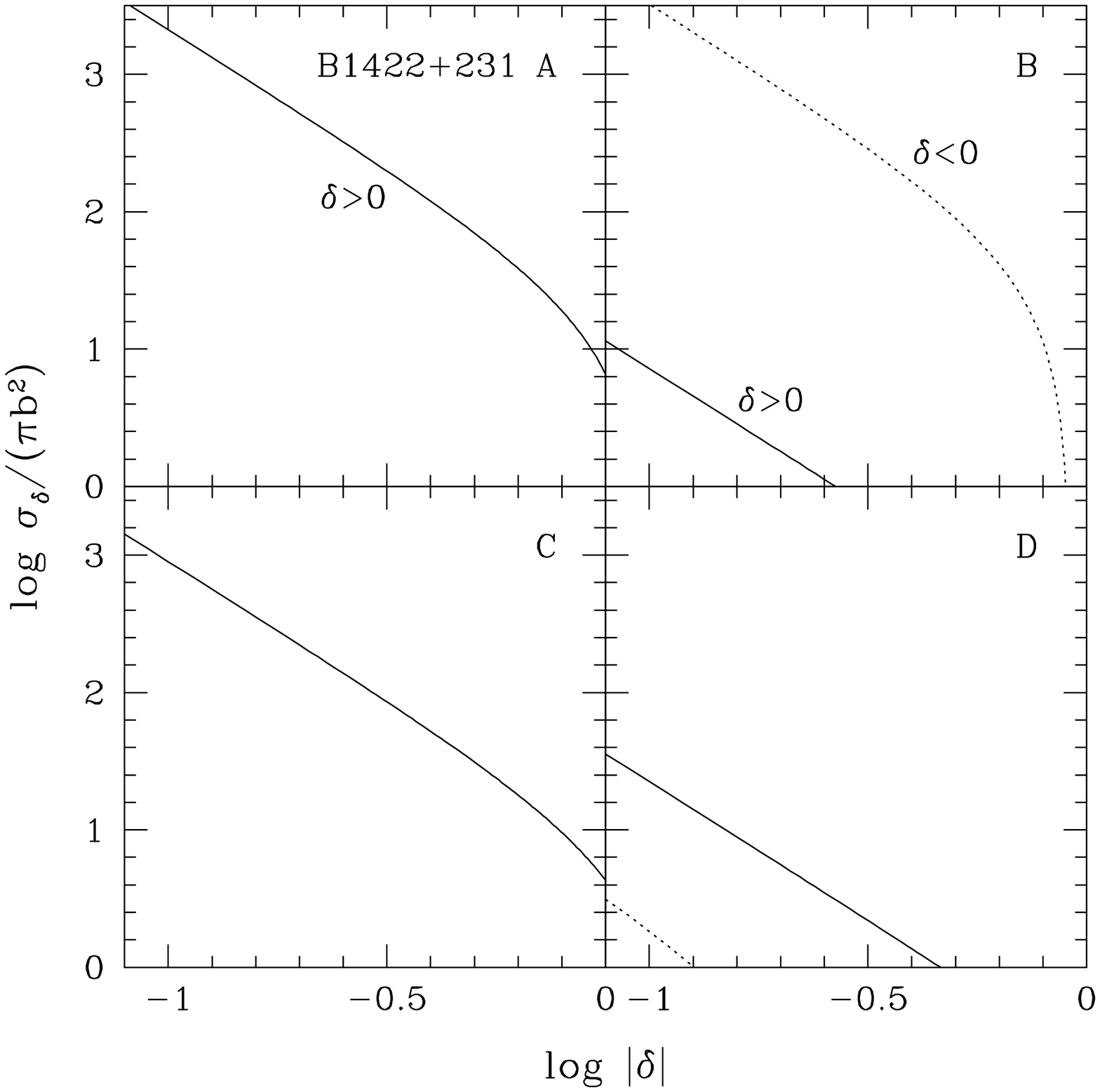}{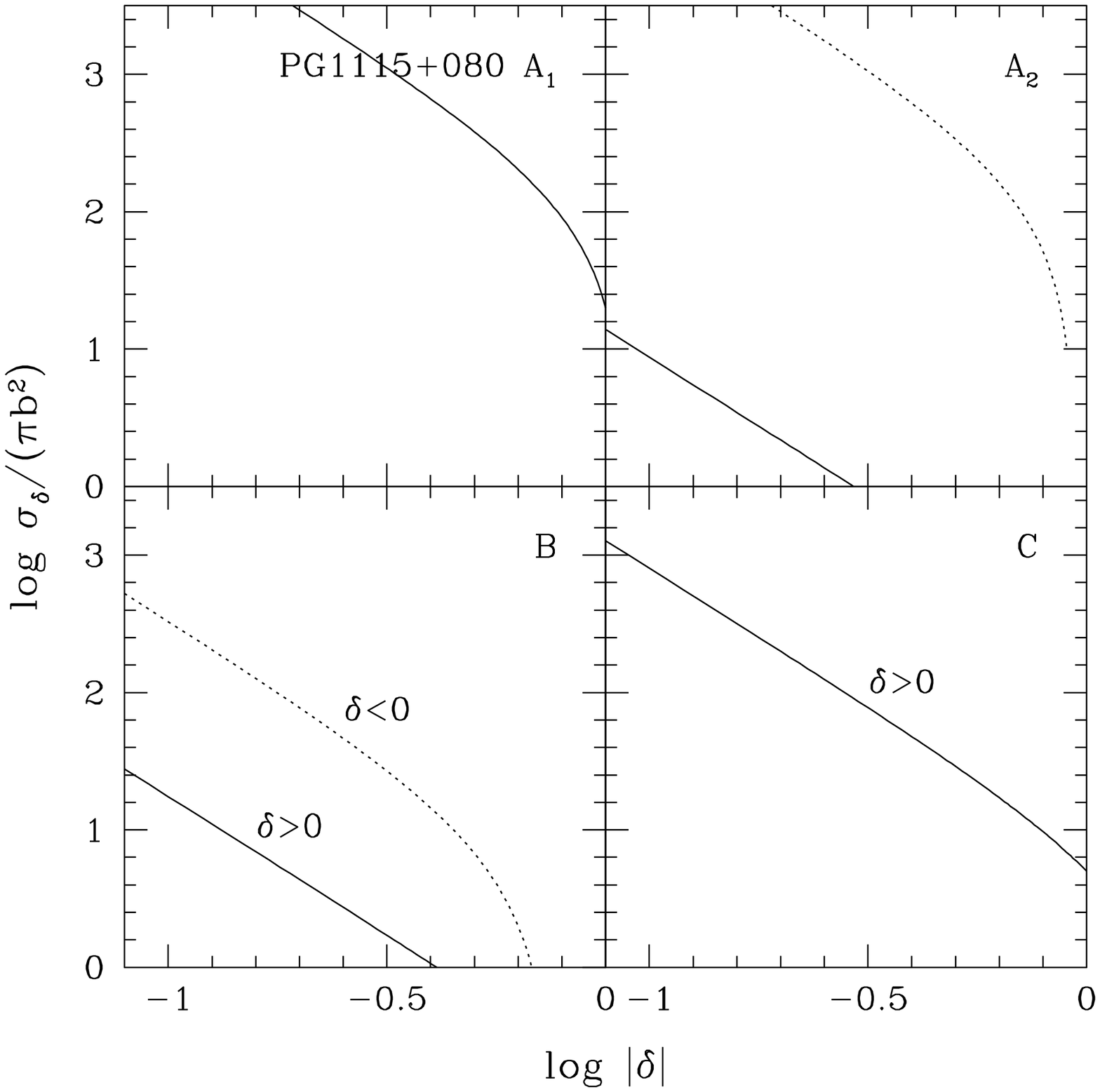}
\caption{
Estimates of the cross sections for magnification perturbations
stronger than $\d$.  Solid curves are for $\d>0$ and dotted curves
are for $\d<0$.  The clumps are assumed to lie in the lens galaxy
($\bt=0$).
{\it (Left)\/} B1422+231.
{\it (Right)\/} PG~1115+080.
}\label{fig:AofD}
\end{figure*}

\reffig{AofD} shows the estimated cross sections for an SIS clump
embedded in the lens galaxy, as a function of $\d$, for each
image in the two systems. In B1422+231 there is a substantial
cross section for the positive-parity images A and C to be made
{\it brighter\/} by substructure ($\d>0$). There is a smaller cross
section for negative-parity image B to be likewise enhanced, but a
much more significant cross section for it to be made {\it dimmer\/}
($\d<0$). The faint negative-parity image D is much less likely to
be affected, and due to its faintness it is more likely to be
brightened than dimmed. The results for PG~1115+080 are similar:
positive-parity images (A$_1$ and C) are always brightened by an
SIS clump, while negative-parity images (A$_2$ and B) can be
brightened but are much more likely to be suppressed. The
differences between positive-parity images and negative-parity
images are similar to those discussed by Schechter \& Wambsganss
(2002). Although those authors were studying microlensing
(perturbations by point-mass stars), the qualitative features
are similar because the conceptual differences between SIS and
point-mass clumps are small.

\begin{figure*}[t]
\plottwo{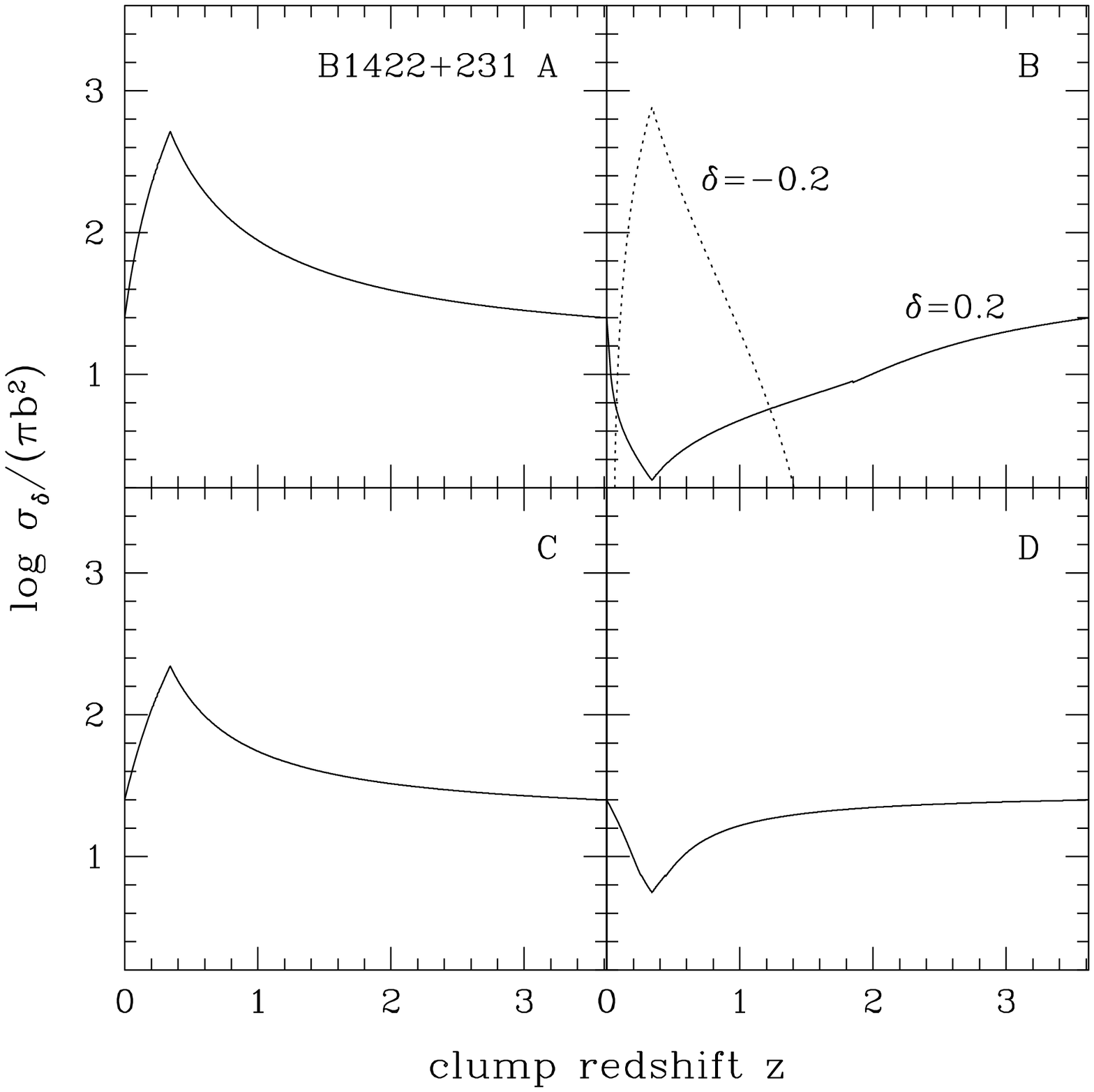}{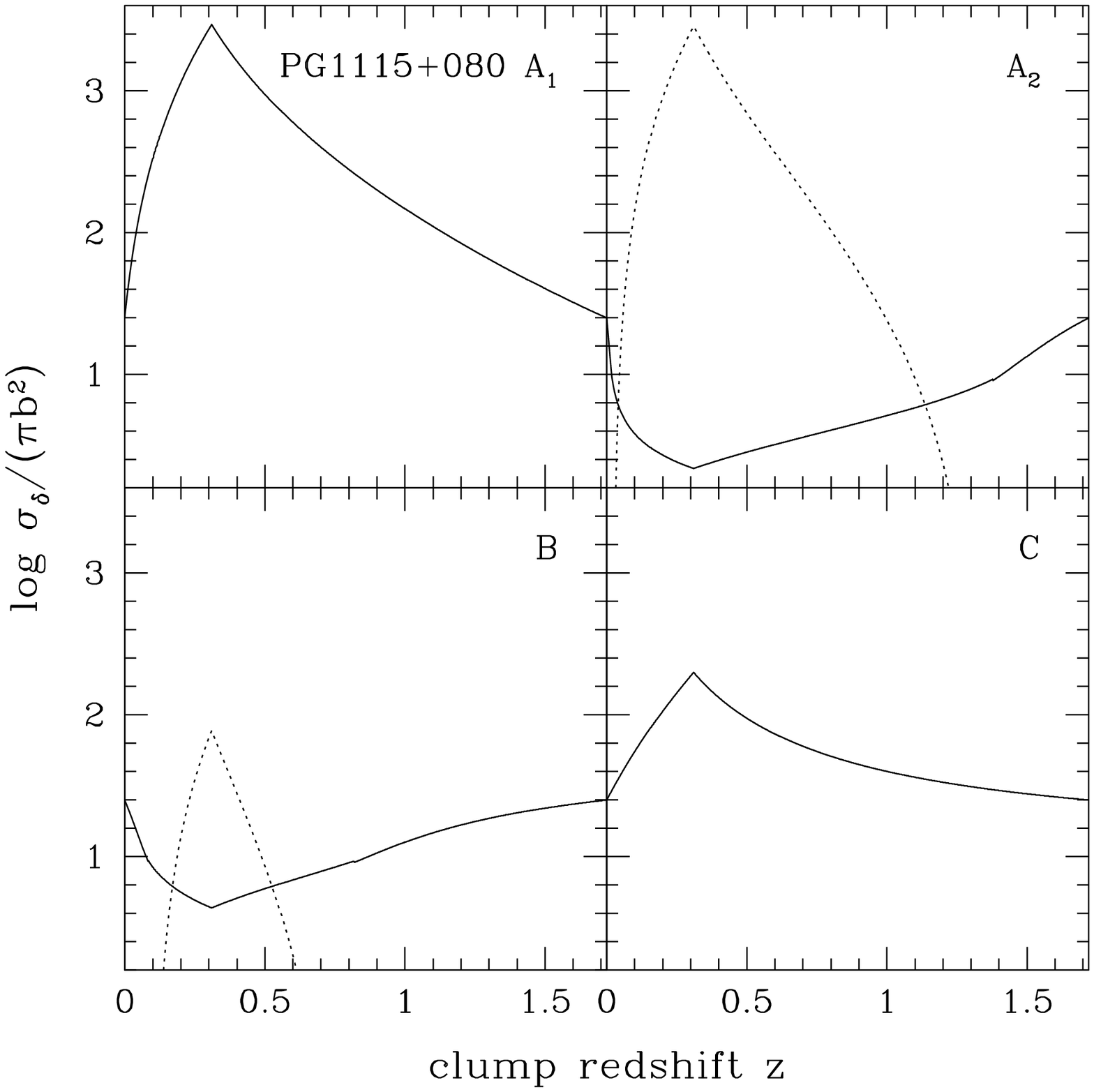}
\caption{
Estimates of the cross sections as a function of the clump
redshift. Solid curves are for $\d=0.2$ and dotted curves are for
$\d=-0.2$. Note that the area is given in units of the Einstein
radius $b$, so effectively it is the Einstein radius (not the clump
mass) that is held fixed as the redshift changes.  The cosmology is
$\Omega_M=0.3$ and $\Omega_\Lambda=0.7$.
{\it (Left)\/} B1422+231.
{\it (Right)\/} PG~1115+080.
}\label{fig:AofZ}
\end{figure*}

\reffig{AofZ} shows how the cross sections change if the clump is
taken out of the lens galaxy and placed elsewhere along the line of
sight. In general the cross sections peak at the redshift of the
lens galaxy, but the curves are relatively broad with a FWHM of
$\Delta z \sim 0.11$--$0.34$. In other words, clumps could differ
in redshift from the lens galaxy by up to several tenths and still
have a significant effect on the images. For positive-parity
images, the cross section for brightening ($\d>0$) even has tails
extending to $z=0$ and to the source redshift. For negative-parity
images, the cross section for dimming ($\d<0$) is broad but falls
to zero at the redshifts where $\bt > (\k+\g)^{-1}$ (see
eq.~\ref{eq:effparity}). The cross section for brightening ($\d>0$)
of negative-parity images is actually a {\it minimum\/} at the lens
redshift and increases moving toward $z=0$ or the source redshift.
Although conceptually interesting, this behavior may not be of
practical importance because the $\d>0$ cross section tends to be
much smaller than the corresponding $\d<0$ cross section for a
negative-parity image.

These results reveal the interesting possibility that clumps at a
range of redshifts can have a substantial effect on lensing. The
clumps need not be associated with the lens galaxies but could be
the sort of isolated small halos predicted to be common in
hierarchical structure formation models.  It is not known whether
isolated clumps are likely to be important in practice. Although
their number per unit volume is expected to be much lower than that
for embedded clumps in overdense regions like galaxies (at least
for CDM), they are viable for lensing over a much larger volume.
Determining the relative importance of the two populations will
involve a detailed calculation combining lensing cross sections
with appropriate clump mass functions (J.\ Chen et al., in
preparation). The question is significant because if isolated
clumps cannot be ignored, they will weaken the argument that
substructure lensing opposes a Warm Dark Matter model (see \S1).

\subsection{Comments}

So far this analysis has considered only a single clump, but it
can be extended to a clump population if we make the assumption
that the optical depth is low enough that the clumps essentially
act independently. In this case, the optical depth due to clumps
in a patch of sky is estimated by summing the cross sections of
all clumps in the patch and dividing by the area of the patch,
\begin{equation}
  \tau(\d;\k,\g) = \int_{0}^{z_s} dz\,D(z)^2\,\frac{dD}{dz}
    \int dM\,\frac{dn}{dM}\,\s(\d;\k,\g,\bt,M)\,,
    \label{eq:tau1}
\end{equation}
where $\s$ is the cross section for a single clump, $dn/dM$ is the
mass function of clumps, and $D(z)$ is the comoving distance. The
optical depth is independent of the area of the patch provided
that the number density of clumps is constant across the patch.
The assumption that the clumps act independently can be checked
{\it a posteriori\/} by seeing whether the estimated optical depth
is small. If so, the assumption is valid. If not, the estimated
optical depth may not be accurate --- but it is still useful,
because it at least indicates that the optical depth is not
vanishingly small. That alone may be sufficient for some questions,
such as the importance of isolated clumps. In any case, a numerical
evaluation of the double integral in \refeq{tau1} is much faster
than a Monte Carlo or ray shooting simulation of lensing with
clumps, so it should be valuable for exploring the optical depth in
large parameter spaces.

Finally, a convenient feature of SIS clumps is that the mass only
affects the Einstein radius $b$, and the cross section scales with
$\pi b^2$.  Thus, the scaled cross section $\hat\s = \s/(\pi b^2)$
is independent of clump mass, so we way write
\begin{equation}
  \tau(\d;\k,\g) = \pi \int_{0}^{z_s} dz\,D(z)^2\,\frac{dD}{dz}\,
    \hat\s(\d;\k,\g,\bt) \int dM\,\frac{dn}{dM}\,b^2 .
\end{equation}
The inner integral is just an integral over the mass function
weighted by $b^2$. For SIS clumps, $b \propto \s^2 \propto M^{2/3}$
(see eq.~\ref{eq:SISb}), so for mass functions steeper than
$dn/dM \propto M^{-4/3}$ the optical depth will be dominated by
low-mass clumps.

\section{Conclusions}

In the context of substructure lensing, the general problem of a mass
clump anywhere along the line of sight to a strong lens can be mapped
onto an equivalent problem of a clump in a simple convergence and shear
field; thus, the theory can easily accommodate clumps at arbitrary
redshifts. If a clump can be approximated as an SIS, the clump plus
external field problem is analytically tractable, yielding simple
formulas for the cross sections relevant for substructure lensing.
Both of these results will make it possible to explore substructure
lensing effects in large parameter spaces quickly and easily.

The analytic results already reveal two interesting features of
substructure lensing. First, the sign of the perturbation depends
on the parity of the original image. SIS clumps always make
positive-parity images brighter, and they usually make
negative-parity images fainter. There is some probability that SIS
clumps can make negative-parity images brighter, but it is usually
much smaller than the probability of dimming, except for clumps far
in redshift from the lens galaxy. The qualitative difference between
positive-parity and negative-parity images has been remarked by
Metcalf \& Madau (2001) for perturbations by galactic subclumps,
and studied in detail by Schechter \& Wambsganss (2002; also Witt,
Mao \& Schechter 1995) for perturbations by stars. The effect is
qualitatively similar for the two types of perturbers, although it
appears to be more dramatic for stars. It is a distinctive feature
of substructure lensing (whether by stars or by larger, extended
clumps) that will be useful for distinguishing the phenomenon from
other systematic effects in lens flux ratios.

The second qualitative result is that a clump need not reside in
the halo of the lens galaxy in order to have a significant lensing
effect. Lensing is sensitive to clumps anywhere along the line of
sight --- not uniformly, but with a redshift dependence that I have
quantified.  The sensitivity function peaks at the lens redshift,
but it has a width of several tenths in redshift and tails that
extend to $z=0$ and to the source redshift.  If CDM is right,
isolated small halos may be much less abundant (per unit volume)
than clumps within galaxy halos, but they are viable over such a
large volume that they may still be relevant for substructure
lensing.  If WDM is right and galaxies are relatively smooth,
isolated clumps could be the dominant cause of substructure lensing.
Using predicted clump populations to evaluate the relative importance
of isolated and embedded clumps is the subject of a work in progress
(J.\ Chen et al., in preparation).  This issue may ultimately
affect how substructure lensing is interpreted, and how strongly
it supports CDM and rules out WDM.

\acknowledgements
I would like to thank Andrey Kravtsov, Jackie Chen, and Daniel Holz
for interesting discussions that stimulated this project.  In
particular, Andrey Kravtsov and Daniel Holz independently suggested
that it might be worthwhile to consider isolated clumps.
This work was supported by NASA through Hubble Fellowship grant
HST-HF-01141.01-A from the Space Telescope Science Institute, which is
operated by the Association of Universities for Research in Astronomy,
Inc., under NASA contract NAS5-26555.

\onecolumn

\end{document}